\documentclass[twocolumn,prd,nofootinbib,showpacs]{revtex4}
\usepackage{graphicx}

\def\som{Sommerfeld enhancement~}
\def\26{{$\sigma_{26}$}}

\newcommand{\be}{\begin{equation}}
\newcommand{\ee}{\end{equation}}
\newcommand{\bea}{\begin{eqnarray}}
\newcommand{\eea}{\end{eqnarray}}
\newcommand{\sigmav}{$\langle\sigma v\rangle$}

\newcommand{\lsim}{\mathrel{\mathop{\kern 0pt \rlap
  {\raise.2ex\hbox{$<$}}}
  \lower.9ex\hbox{\kern-.190em $\sim$}}}
\newcommand{\gsim}{\mathrel{\mathop{\kern 0pt \rlap
  {\raise.2ex\hbox{$>$}}}
  \lower.9ex\hbox{\kern-.190em $\sim$}}}

\begin{document}
\title{Conservative upper limits on WIMP annihilation cross section from Fermi-LAT $\gamma$-rays}
\author{Francesca Calore}\email{francesca.calore@desy.de}
\affiliation{ II. Institute for Theoretical Physics, University of Hamburg, Luruper Chaussee 149, 22761 Hamburg, Germany}
\author{Valentina De Romeri}\email{deromeri@ific.uv.es}
\affiliation{Astroparticle and High Energy Physics Group, IFIC (CSIC - Universidad 
de Valencia) - Edificio Institutos de Investigacion, C/ Catedratico 
Jose Beltran 2, E-46980 Paterna (Valencia), Spain}
\author{Fiorenza Donato}\email{donato@to.infn.it}
\affiliation{Dipartimento di Fisica Teorica, Universit\`a di Torino
and INFN, Via P. Giuria 1, 10122 Torino, Italy}

%\date{\today}

\begin{abstract}
The spectrum of an isotropic extragalactic $\gamma$-ray background (EGB)  
has been measured by the Fermi-LAT telescope at high latitudes.
Two new models for the EGB are derived from the subtraction of
 unresolved point sources and extragalactic 
diffuse processes, which could explain from 30\% to 70\% of the Fermi-LAT 
EGB.  
Within the hypothesis that the two residual EGBs are entirely due to the  
annihilation of dark matter (DM) particles in the Galactic halo, we   
obtain stringent upper limits on their annihilation cross section. 
Severe bounds on a possible Sommerfeld enhancement of 
the  annihilation cross section are set as well. 
Finally, we consider models for DM  annihilation  depending on the  
inverse of the velocity and associate the EGBs to photons arising 
from the annihilation of DM in primordial halos.
Given our choices for the EGB and the minimal DM modelling, 
the derived upper bounds are claimed to be $conservative$.

\end{abstract}
\pacs{95.30.Cq,95.35+d,95.85.Pw,96.50.sb}
\maketitle

\section{Introduction}

The indirect search for dark matter (DM) through its annihilation
products in rare charged cosmic rays (CRs) and in multi-wavelength
channels requires  very accurate measurements and an unambiguous 
estimation of all the possible backgrounds to the DM signal. 
In the last years, dedicated experiments have provided 
unprecedented results by extending the energy ranges of the measured 
cosmic species as well as the precision of the data
\cite{2009Natur.458..607A,2010APh....34....1A, 2009PhRvL.102r1101A,
2009A&A...508..561A, 2009PhRvL.102e1101A,Adriani:2010rc,LATIsotropicSpectrum}.
 Further data are expected by the Fermi-LAT 
and Pamela on-going missions, and by the AMS-02 experiment 
on board the International Space Station.
From the theoretical side, many efforts have been addressed to a better 
and increasingly detailed modellization of the astrophysical 
processes which shape, at different levels, the observed fluxes. 
%%% pbar
Data from cosmic antiprotons \cite{2009PhRvL.102e1101A,Adriani:2010rc}
have been shown to be compatible with 
the standard production from CRs impinging on the interstellar gas 
\cite{2009PhRvL.102g1301D}. 
%%% positrons
The anomalous increasing positron fraction 
measured by Pamela \cite{2009Natur.458..607A,2010APh....34....1A}
and confirmed by Fermi-LAT \cite{positron_fermi} may be explained 
by emission from near pulsars over-imposed to a standard CR
population   \cite{2010A&A...524A..51D,2009JCAP...01..025H}. 
Alternatively, a DM component with very high cross section or
sources concentration has been invoked \cite{2008NuPhB.800..204C,2008PhRvD..78j3520B,2009PhRvD..79a5014A, Cholis}.
%%% gamma rays
Unprecedented $\gamma$-ray measurements by Fermi-LAT
have boosted interpretation of diffused and point sources
emission in terms of exotic components from DM annihilation 
in the halo of the Milky Way, in extragalactic near objects  or 
in cosmological structures \cite{LATIsotropicSpectrum,2010ApJ...712..147A,2010JCAP...05..025A,2010JCAP...04..014A}. 
The very signature would be the monochromatic line, 
which nevertheless provides tiny signal on a remarkable background \cite{2010PhRvL.104i1302A}. 
\\
The high latitude  $\gamma$-ray emission measured by Fermi-LAT \cite{LATIsotropicSpectrum},
given its reduced contamination by galactic sources, 
can be  a powerful tool to set limits on the contribution of DM to the measured 
flux. The data are indeed the result of a non trivial 
subtraction procedure and  
show a high isotropic feature.
\\
The  aim of the present research is to set $conservative$
upper limits on the galactic weakly interacting massive particle (WIMP) DM annihilation cross section into $\gamma$-rays.
Several upper limits have been obtained 
through different and complementary indirect research means
\cite{Cirelli:2009dv,2011PhRvD..84b7302G,2009PhRvL.102g1301D,
2010JCAP...04..014A,Baxter:2011rc,Siffert:2010cc,Abdo:2010ex,
hess_dwarfs,Abramowski:2011hc,fermi_dwarfs_2011,koushiappas}. 
However, it is usually not straightforward to compare
these results, given the model dependence, the different assumptions
on the astrophysical backgrounds, and the theoretical uncertainties.
We will confront the $\gamma$-rays coming 
both from the DM halo and high-redshift protohalos with the background observed by Fermi-LAT at high latitudes. The conservative approach is
achieved - in addition to prudent assumptions on the particle physics model and DM distribution in the Galaxy - through the comparison
of the putative DM signal with a high latitude diffuse emission spectrum (i.e. EGB) obtained with minimal subtractions of known unresolved sources. 
\\ 
Our paper proceeds as follows. 
In Sect. 2 we discuss the possible contributions to the 
high latitude $\gamma$-ray emission from unresolved point sources
and truly diffuse processes. We subtract the non-negligible fluxes 
to the Fermi-LAT data and draw two possible scenarios for the 
high latitude emission. 
In Sect. 3 we derive conservative upper limits to the DM annihilation cross 
section by identifying the residual $\gamma$-ray flux with $\gamma$-rays 
from DM annihilation in the galactic halo and in primordial DM small halos at high redshift.
In the latter case, we study models in which 
the DM annihilation cross section has an explicit dependence 
on the inverse of the velocity. 
We discuss also a possible Sommerfeld enhancement of the annihilation cross section
and derive limits on its amplitude. 
In Sect. 4 we draw our conclusions.

\section{The extragalactic $\gamma$-ray background} 
\label{EGB} 
A diffuse $\gamma$-ray emission has been measured by the Fermi-LAT detector at high
latitudes ($|b|>10^{\rm o}$) \cite{LATIsotropicSpectrum}. 
The spectrum has been obtained after the subtraction from the data of 
 the sources resolved by the telescope, the (indeed model dependent) 
 diffuse galactic emission, 
the CR background in the detector and the solar $\gamma$-ray emission. The resulting flux 
 decreases with a power law of the photon energy with spectral index $2.41\pm0.05$.  It shows a highly isotropic sky distribution and is
generically classified as an extragalactic $\gamma$-ray background (EGB). 
\\
%% Introduciamo le unresolved point sources.
The 1451 sources listed in the First Fermi-LAT catalog (1FGL) \cite{1FGL}
represent the best-resolved survey of the sky in the 100 MeV to 100 GeV energy range. 
%Out of the 1FGL 1451 sources, 821 (56$\%$)
%have been associated with at least one non-$\gamma$-ray counterpart. The largest population is constituted by Active Galactic Nuclei (AGNs), mainly blazar candidates, with 685 associations.
For each low--flux source there may be a large number of \textsl{unresolved} 
point sources which have not been detected because of selection
effects, 
%(e.g. the local background was too large or the photon index was too soft, or a combination of both), 
or too low emission. 
%Ref. \cite{2010ApJ...720..435A} presents a study of the high latitude sources which have not been detected by Fermi-LAT 
%because of these selection effects, but have a flux which is formally larger than the faintest detected source. 
%The maximum contribution (arising from the
%integration of the logN-logS distribution down to zero flux) of unresolved sources to the diffuse background is estimated  to be $2.39(\pm
%0.48) \cdot 10^{-6}$ ph cm$^{-2}$ s$^{-1}$ sr$^{-1}$, 
%which represents $23(\pm 5)\%$ of the Fermi-LAT high latitude isotropic diffuse background \cite{LATIsotropicSpectrum}.
%\\
%%% Contributi NON TRASCURABILI all'EGB 
%Several possible sources to the residual EGB have been explored in the literature. From the analysis in Ref. \cite{2010ApJ...720..435A}, it is likely 
%It is shown that 
Most of the unassociated high latitude sources are blazars,  a class of Active Galactic Nuclei (AGNs), and their pile to the EGB with the largest flux \cite{2010ApJ...720..435A}. 
Galactic resolved pulsars and Milli-Second Pulsars (MSPs) represent the second largest population  in the Fermi-LAT catalog
\cite{1FGL,2010ApJS..187..460A} and they are expected to contribute significantly to the putative EGB.
A non-negligible $\gamma$-ray flux seems to be guaranteed by unresolved normal star-forming galaxies \cite{EGBStarformgal2010}. 
Ultra-high energy CRs (UHECRs) may induce secondary electromagnetic cascades, originating neutrinos
and $\gamma$-rays at Fermi-LAT energies \cite{EGBCMB1}.
%All of these possible bricks to the residual EGB will be discussed in some detail in the following of this Section. 
%Only 
Contributions from unresolved blazars and MSPs are believed to contribute at least few percent to the Fermi-LAT 
EGB, while predictions for star-forming galaxies and UHECRs are highly model dependent. 
\\
%% Contributi  TRASCURABILI all'EGB 
Other astrophysical sources may emit in the 
high latitude $\gamma$-ray sky:
i) radio-quiet AGN \cite{RadioQuietPhys, EGBInoue}, and 
Fanaroff and Riley radio galaxies of type I and II
\cite{2010ApJ...720..912A,EGBradio,2011arXiv1103.3946I}
whose contribution is strongly model dependent 
and likely bound to few percent of the EGB;
ii)  $\gamma$-ray bursts (GRBs), estimated less than 1\% of the diffuse extragalactic 
$\gamma$-ray background \cite{GRBs};
iii) star-burst and luminous infrared galaxies.  
The relevant flux may cover a significant fraction of the
EGB ($\leq  20\% $) \cite{EGBStarburst}, but the model dependence is such 
to prevent firm statements on the relevance of this extragalactic source;
iv)  nearby clusters of galaxies, which  could yield about $1\% - 10\%$ 
of the EGRET EGB  \cite{EGBcluster,Dermer2003,Pfrommer2008};
v) gravitational induced shock waves, produced during cluster mergers and large-scale structure formation, whose fluxes are quite model dependent  and may reach few percent
\cite{EGBLSS2002,EGBLSS2003}.
All these $\gamma$-ray sources 
have been shown to contribute to less 
than  1\% of the Fermi-LAT EGB  or to be too highly model dependent. In the latter case, a very high uncertainty band would be associated with the $\gamma$-ray source, whose lower limit likely gives a negligible contribution to the Fermi-LAT EGB. They will therefore 
 be neglected in rest of our paper. 
\\ 
%%Declaratoria sul metodo
In the following, we describe few classes of $\gamma$-ray emitters whose 
unresolved flux is firmly estimated in a non-negligible Fermi-LAT EGB percentage. 
In a conservative scenario (Model I), we will subtract AGN and MSPs
 to the Fermi-LAT EGB as derived in Ref. \cite{LATIsotropicSpectrum}. 
A more relaxed model (Model II) will be drawn by the further subtraction 
of a minimal flux from star-forming galaxies and CRs at the highest 
energies.

%%%%%%%%%%%%%%%%%%%%%%%%%%%%%%%%%%%%%%%%%%%%%%%%%%%%%%%%%%%%%%%%%%%%%%%%%%%%%%%%%%%%5
\subsubsection{BL Lacs and FSRQs} 
 \label{sec:blazars}  
% An AGN is a compact region at the center of a
%galaxy, probably originated by galactic matter accretion onto a super-massive black hole. The released large amount of gravitational energy
%flows away through powerful jets of relativistic particles which in turn produce X and $\gamma$ radiation. 
%Blazars are those AGNs for which the jets are close to the l.o.s. The blazars classification includes BL Lacs,
%which present a complete or nearly complete lack of emission lines, and FSRQs.  
Blazars constitute the 
class of $\gamma$-ray emitters with the largest number of identified members. 
Therefore, unresolved blazars are expected to have a sizable
contribution to the EGB,  \cite{EGRET}. The largest uncertainties in determining the blazars
contribution are their unknown spectral energy distribution
and luminosity function \cite{EGBInoue,Stecker:2010di}. 
In addition to phenomenological predictions,
%Theoretical estimations of the $\gamma$-ray flux from unresolved FSRQs are 
%derived in Ref. \cite{Stecker:2010di}. Adopting a
%power law $\gamma$-ray energy spectrum and assuming that the
%$\gamma$-ray luminosity of an FSRQ is, on average, proportional to its radio luminosity, 
%the authors find that a scenario in which the EGB is dominated by emission from unresolved blazars
%is compatible with the Fermi source count data. 
%%
%In Ref. \cite{EGBInoue} it was shown that the EGB flux could 
%be composed of blazars and non-blazar AGNs in the luminosity-dependent density evolution (LDDE) 
%SED blazar model, consistently with the EGRET blazar data and the X-ray AGN surveys. 
%Different models for the $\gamma$-ray luminosity function have been considered, as well different model parameters 
%for the non-blazars $\gamma$-ray emission, leading to different theoretical predictions
%\\
%In addition to phenomenological predictions largely affected by systematic and theoretical uncertainties, 
an analysis of the observed source count
distribution through Monte Carlo simulations has been performed in Ref. 
 \cite{2010ApJ...720..435A}.
%can determine the EGB contribution of an unresolved source class.
%In Ref.  \cite{2010ApJ...720..435A} the source count distribution of blazar objects in a %simulated sample has been performed. 
The reliability of the algorithm relies onto a good agreement with the real data,
from the comparison of reconstructed $\gamma$-ray fluxes and spectral
properties of the sources. 
%Notably, the number of identified simulated sources does not exceed the observed sample.
%The simulated contribution of unresolved FSRQs and BL Lacs to the EGB  is reported in Fig. 20 
%of Ref. \cite{2010ApJ...720..435A}. It is obtained by
%integrating the simulated unresolved source count distribution from the flux of the faintest source in their sample ($S_{min}^{BL Lac} = 9.36
%\cdot 10^{-10} {\rm ph \;\ cm^{-2} \; s^{-1}} $ and $S_{min}^{FSRQ} = 1.11 \cdot 10^{-8}
% {\rm ph \; cm^{-2} \; s^{-1}}$) to a maximal flux intensity
%$S_{max} = 10^{-3} {\rm ph \;\ cm^{-2}\; s^{-1}}$.
The energy spectrum is well described by a power-law for both 
FSRQs (softer) and BL Lacs one (harder), 
being the intersection between the two fluxes at about 400 MeV.
Following our conservative approach - which is meant to consider the minimum unavoidable contribution to the EGB from unresolved 
astrophysical sources - we
will adopt blazar contributions from 
 the curves delimiting the lower uncertainty bands displayed in Fig. 20 of Ref. \cite{2010ApJ...720..435A}. 
The ensuing flux is  displayed in our Fig. \ref{fig:EGB_Model_I} as dotted (dot-dashed) line for BL Lacs (FSRQs) contribution.

%%%%%%%%%%%%%%%%%%%%%%%%%%%%%%%%%%%%%%%%%%%%%%%%%%%%%%%%%%%%%%%%%%%%%%%%%%%%%% 
\subsubsection{Pulsars and MSPs}
\label{sec:PulsarsAndMSPs}
As a result of their short periods, typical MSPs may be brighter in the $\gamma$-rays and much older than ordinary pulsars \cite{EGBPulsar2009}.  
The ages of MSPs generally exceed the oscillation time across the 
galactic disk by a large factor so that MSPs are expected to be more prevalent at
high latitudes. On the contrary young, energetic ordinary pulsars are more concentrated close to the Galactic plane, where they were born. 
 In the first year of Fermi-LAT observations \cite{1FGL}, 63 pulsars have been identified. Among them: 
(i) 16 pulsars at $|b| > 10^{\rm o}$, of which 11 are MSPs; (ii) 5 MSPs at $|b| > 40^{\rm o}$ and (iii) 1 MSP at $|b| > 60^{\rm o}$.   
%The empirical analysis conducted in
%Ref. \cite{EGBPulsar2009} finds  the $\gamma$-ray flux distribution of MSPs (i.e. the logN-logS distribution)  to be close to Euclidean (i.e.
%homogeneous spatial distribution in a three-dimensional space) at the high end of the logN-logS flux distribution,  in contrast to the
%ordinary $\gamma$-ray pulsars which are essentially found in the galactic disk. 
%%
%\\
We estimate a minimal but not negligible contribution of the unresolved MSPs population to the $\gamma$-ray flux at high latitudes. 
We adopt an empirical prescription outlined 
in Ref. \cite{2010arXiv1011.5501S}, 
which is based on the spectra of the eight MSPs detected by Fermi in the first 9 months \cite{PulsarScience} of operation.  
The differential energy spectra of the Fermi-detected MSPs are well described by a truncated power law: 
\begin{equation} \frac{dN}{dE}
= K E^{-\Gamma}e^{-E/E_{cut}}. 
\label{MSPs_spectrum} 
\end{equation}  
$\Gamma$ and $E_{cut}$ are assumed to be  $\left\langle \Gamma \right\rangle= 1.5$ and $\left\langle E_{cut}\right\rangle = 1.9$ GeV, while $K$ has been obtained for $|b| \geq 40^{\rm o}$. 
\\
In order to evaluate Eq. (\ref{MSPs_spectrum}) for different observational regions - namely changing the normalization $K$ - we follow the
prescriptions given in Ref. \cite{EGBPulsar2009}. Assuming a disk-like latitude profile, the ratio of the average intensities at different
latitudes is given by:  
\begin{equation}
 \frac{I_{MSP}(|b| \geq b_{1})}{I_{MSP}(|b| \geq b_{2})} =
\frac{\rm{ln[(sin|b_{1}|)^{-1}]}}{\rm{ln[(sin|b_{2}|)^{-1}]}},
\label{lat_profile} 
\end{equation} 
where $I_{MSP}(|b| \geq b_{i})$, $i = 1, 2$, is the
average MSP intensity over a solid angle $\Omega = 4 \pi (1 - {\rm sin}|b_{i}|)$ defined by the integration from the minimal latitude $b_{i}$ up to $90^{\rm o}$, written as: 
\begin{equation} I_{MSP} \equiv 
\frac{S_{tot}}{\Omega} =  
\frac{S_{min}}{\Omega} \cdot \left(\frac{\delta - 1}{\delta - 2}\right) \cdot \left(\frac{S_{min}}{S_{th}}\right)^{1 - \delta}
\cdot N(S>S_{th}). 
\label{I_MSP} 
\end{equation} 
$S_{min}$ refers to the assumed Euclidean logN-logS flux distribution of the galactic
MSP population, which is parametrized by a power-law with spectral index $\delta = 2.5$ for $ S \geq  S_{min}$. According to
\cite{EGBPulsar2009}, we set $S_{min} = 10^{-10}$ ph s$^{-1}$ cm$^{-2}$. 
$N(> S_{th})$ is the number of resolved sources above a given flux threshold $S_{th}$.
 We update the estimation for $I_{MSP}$ in Ref. \cite{EGBPulsar2009} with the more recent observations for $|b| \geq
10^{\rm o}$ reported in Ref. \cite{PulsarScience}, where 8 MSPs have been found above 
$S_{th}= 2 \cdot 10^{-8}$ ph s$^{-1}$ cm$^{-2}$ (lowest detected MSP flux).  We find: 
\begin{equation} 
I_{MSP}(|b| \geq 10^{\rm o}, E > 100 ~ {\rm MeV}) = 6.54 \cdot 10^{-7}  {\rm cm^{-2} s^{-1}sr^{-1}}.
\end{equation}
$K$  is then derived from: 
\begin{equation} 
I_{MSP} = \int ^{E_{max}}_{E_{min}} \frac{dN}{dE} dE \ , 
\label{normalization} 
\end{equation}
where $\frac{dN}{dE}$ refers to Eq. (\ref{MSPs_spectrum}).
Cross-checking the average MSP intensities obtained with the 
prescription outlined above, and the results in Ref. \cite{2010arXiv1011.5501S} for $|b| \geq 40^{\rm o}$, 
we find a relative difference of about $30\%$, due to the theoretical uncertainties on the assumed logN-logS 
and the latitude profile. 
We consider such a discrepancy as an empirical theoretical uncertainty on the determination of $K$, 
and fix the unresolved MSPs contribution subtracting a $30\%$ uncertainty from the estimated 
average intensity. The MSP contribution is shown in Fig. \ref{fig:EGB_Model_I} as a double dot-dashed line.

%%%%%%%%%%%%%%%%%%%%%%%%%%%%%%%%%%%%%%%%%%%%%%%%%%%%%%%%%%%%%%%%%%%%%%%%%%%%%% 
\subsubsection{Star-forming Galaxies}
\label{sec:StarformingGalaxies} 
Unresolved normal star-forming galaxies  are expected to give a guaranteed contribution to the
high latitude isotropic diffuse $\gamma$-ray background.  Fermi-LAT
 has identified the source of the diffuse emission from our Galaxy
due to the collisions of CRs with interstellar gas, leading to $\gamma$-rays from $\pi^{0}$ decay in flight. 
This observation provides a ground to estimate the  $\gamma$-ray luminosity of star-forming galaxies,  by scaling 
the CR flux with the massive star formation rate and fixing the amount of the gas in the external galaxy.  
%The lack of statistics does not allow a source count population study for this source class. Several models 
% rely on the determination of the $\gamma$-ray luminosity function describing the evolution of space density 
%and luminosity with the redshift 
Theoretical predictions are greatly
affected by uncertainties in the determination of the star formation rate of the galaxies and their gas content \cite{EGBStarformgal2010, Stecker:2010di, EGBStarformgalTotani}. 
 Given the uncertainty surrounding key elements of the determination of this
contribution,  in our strictly conservative approach we do not take into account this component.
%We extend our analysis adopting a more relaxed perspective where also star-forming galaxies contribute. 
In a more relaxed perspective, we consider the lowest predicted contribution from star-forming galaxies \cite{EGBStarformgal2010}. It is derived assuming an increase in the number of star forming galaxies with the redshift. The adopted emission corresponds to the long dashed curve in Fig. \ref{fig:EGB_Model_II}.

%%%%%%%%%%%%%%%%%%%%%%%%%%%%%%%%%%%%%%%%%%%%%%%%%%%%%%%%%%%%%%%%%%%%%%%%%%%%%%%%%%%%
\subsubsection{UHECRs} 
\label{sec:UHECRs} 
UHECRs accelerated in astrophysical objects produce secondary electromagnetic
cascades during their propagation in the cosmic microwave and infrared backgrounds.
Ref. \cite{EGBCMB1} shows that if the primary CRs are dominated by protons, 
such cascades can contribute between 1$\%$ and
50$\%$ of the GeV-TeV diffuse photon flux measured by the EGRET experiment.
In Ref. \cite{EGBCMB}, the EGB spectrum from UHECRs (normalized to the HiRes data) has been obtained through a Monte Carlo simulation of the cascade development and compared with the measurement of the EGB by Fermi-LAT. 
 %Assuming an injection spectrum in the form $dN/dE \propto E^{-\alpha_{g}}$ and
%a homogeneous source distribution up to a maximal redshift, UHE accelerated protons are 
%propagated through the extragalactic space until their energy is below the threshold for $e^{+}e^{-}$ pair production, $E_{min} \simeq
%10^{18}$ eV, or until they reach the Earth. The simulation is performed for two UHECR models, corresponding to a different exponent of the
%generation spectrum: i) $\alpha_{g} = 2.6$ 
%gives the cascade flux for the non-evolutionary dip model with maximal energy of acceleration $E_{max}
%= 1\cdot10^{21}$ eV and maximal redshift $z_{max} = 2$, while ii)   $\alpha_{g} = 2.0$ represents the ankle model with a
%transition from galactic to extragalactic CRs at $5\cdot10^{18}$ eV, for the same values of $E_{max}$ and $z_{max}$. 
In our more relaxed, whether conservative scenario, we 
will subtract the ankle model contribution to the Fermi-LAT EGB \cite{EGBCMB}, which we show in Fig. \ref{fig:EGB_Model_II}. This $\gamma$-ray component has the peculiar behaviour to slightly 
increase with increasing energy, and at 100 GeV may account 8\% of the Fermi-LAT measured EGB.

%%%%%%%%%%%%%%%%%%%%%%%%%%%%%%%%%%%%%%%%%%%%%%%%%%%%%%%%%%%%%%%%%%%%%%%%%%%%%
 %%%%%%%%%%%%%%%%%%%%%%%%%%%%%%%%%%%%%%%%%%%%%%%%%%%%%%%%%%%%%%%%%%%%%%%%%%%%

\begin{figure}[!h] \centering
\includegraphics[scale=0.5]{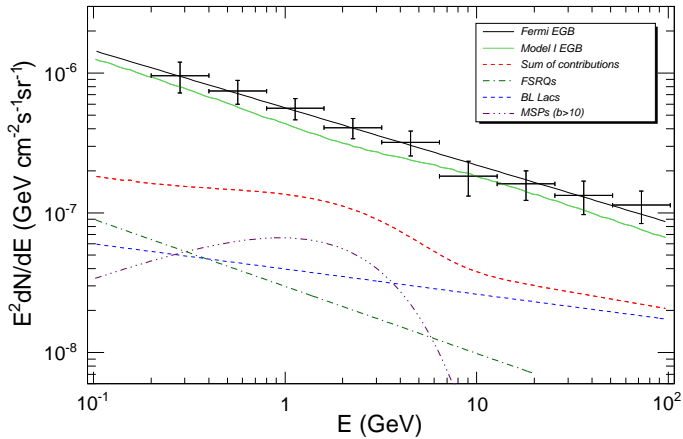}  
\caption{$\gamma$-ray spectrum for $|b|>10^{\rm o}$ latitudes. Fermi-LAT data points 
are displayed along with their power--law fit (solid black curve)
\cite{LATIsotropicSpectrum}. The dotted (blue),  dot-dashed (green) 
and double dot-dashed (purple) curves correspond to BL Lacs, FSRQs 
and MSPs contribution, respectively. The dashed (red) curve is the sum 
of the previous three fluxes. The solid (lower, green) curve is derived 
by subtracting the three contributions to the Fermi-LAT result (Model I).}
\label{fig:EGB_Model_I}  
\end{figure}
\begin{figure}[!h] \centering
\includegraphics[scale=0.5]{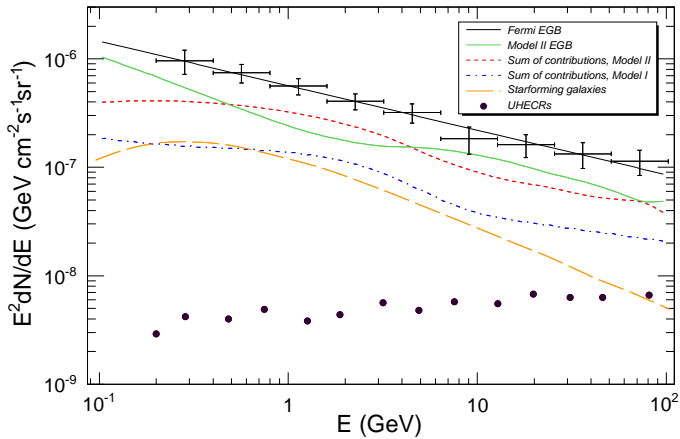}  
\caption{$\gamma$-ray spectrum for $|b|>10^{\rm o}$ latitudes. Fermi-LAT data points 
are displayed along with their power--law fit (solid black curve) \cite{LATIsotropicSpectrum}. 
Dots and long dashed-curve (light blue) correspond to the UHECRs  and star-forming galaxies $\gamma$-ray
fluxes, respectively. The short-dashed (red) curve corresponds to the sum of 
BL Lacs, FSRQs and MSPs contribution (see Fig. \ref{fig:EGB_Model_I}), the short-dashed (blue)
to the sum of the previous components with the star-forming galaxies and UHECRs ones.
The solid (lower, green) curve is derived 
by subtracting all the contributions to the Fermi-LAT result (Model II). } 
\label{fig:EGB_Model_II}  
\end{figure}

%%%%%%%%%%%%%%%%%%%%%%%%%%%%%%%%%%%%%%%%%%%%%%%%%%%%
\subsection{Models for the EGB} 
\label{sec_residui}
As a result of the previous analysis, we now proceed by subtracting
from the Fermi-LAT EGB \cite{LATIsotropicSpectrum} additional contributions 
from unresolved sources at latitudes $|b|>10^{\rm o}$.
The contributions to the EGB that we will remove from the Fermi-LAT spectrum
are minimal. In fact, as extensively explained in the 
previous sections, the predictions that we will take into account
for the relevant unresolved sources are the lowest 
ones according to the literature. In addition, for MSPs we have 
lowered existing 
calculations by updating them to the  Fermi-LAT observations.
\\
In what we label Model I, we subtract 
from the Fermi-LAT EGB \cite{LATIsotropicSpectrum}
the unresolved contributions for both BL Lacs and FSRQs 
as outlined in Sect. \ref{sec:blazars}, and the unresolved MSPs flux obtained according 
to the prescription Sect. \ref{sec:PulsarsAndMSPs}. 
The results are shown in Fig. \ref{fig:EGB_Model_I}, where the Fermi-LAT EGB data
\cite{LATIsotropicSpectrum} 
are shown along with our power--law fit. 
The contributions from BL Lacs, FSRQs and MSPs are identified by dotted, dot-dashed 
and double dotted-dashed curves, respectively. 
The fluxes from the blazar populations follow power--laws, with softer (harder) spectrum for the FSRQs 
(BL Lacs). The crossing point for the two curves is around 300 MeV: above this energy BL Lacs 
flux dominates over the FSRQs one. 
The $\gamma$-rays from unresolved MSPs show a peculiar spectrum peaked at about 1 GeV and dominate
over the blazar spectra from 300 MeV up to 3-4 GeV. 
The sum of the three contributions reflects the MSPs flux shape with a mild bump. 
At about 100 MeV the three sources explain 10\% of the Fermi-LAT EGB and 30\% above  1 GeV. 
The residual flux Model I, 
obtained by subtracting the sum of the three contributions (dashed curve) to 
the Fermi-LAT best fit flux is identified by the lower solid curve.
It is not a net power law due to the dip in the GeV region introduced by the MSPs flux. 
\\
Fig. \ref{fig:EGB_Model_II} refers to the scenario where the 
additional contributions from star-forming galaxies (long dashed line) and UHECRs (solid points) 
as outlined in Sect. \ref{sec:StarformingGalaxies} and Sect. \ref{sec:UHECRs} add to explaining the Fermi-LAT EGB.  
 These two further contributions add with the previous ones 
(blazars and MSPs of Model I) and the total sum is displayed 
by the dashed red line.
The solid (green) curve derives from the subtraction of all these contributions from
 the Fermi-LAT EGB (solid black line fitting the data points) and is labelled Model II hereafter. 
Notably, the contribution from star-forming galaxies turns out to be relevant for $E \leq 1$ GeV, 
whereas the $\gamma$-rays from UHECRs give non-negligible fluxes only at the high-end of
the energy spectrum. We notice that at 100 MeV 
Model II explains about $70\%$ of the Fermi-LAT EGB,  
while above  1-2 GeV they count about $50\%$ of the total. 
To consider additional astrophysical components to the EGB further decreases the residual 
flux (lower solid line) with respect to the Fermi-LAT EGB (upper solid line)
and shrinks the room left to  potential exotic sources, like DM annihilations.

%%%%%%%%%%%%%%%%%%%%%%%%%%%%%%%%%%%%%%%%%%%%%%%%%%%%%%%%%%%%%%%%%%%%%%%%%%%%%
%%%%%%%%%%%%%%%%%%%%%%%%%%%%%%%%%%%%%%%%%%%%%%%%%%%%%%%%%%%%%%%%%%%%%%%%%%%%%
%%%%%%%%%%%%%%%%%%%%%%%%%%%%%%%%%%%%%%%%%%%%%%%%%%%%%%%%%%%%%
\section{Upper bounds on DM annihilation cross section} 
\label{results} 
In this Section we derive conservative upper limits on the 
WIMP annihilation cross section. 
We make the hypothesis that the residual fluxes we have derived in Sect. \ref{sec_residui}
are entirely provided by the $\gamma$-rays produced by 
thermalized WIMP DM in the halo of the Milky Way.

\subsection{$\gamma$-rays from DM annihilation}
\label{Sect:gammaflux}
The flux of $\gamma$-rays $\Phi_\gamma(E_\gamma, \psi)$ originated from
WIMP pair annihilation in the galactic halo
\cite{1990NuPhB.346..129B,1998APh.....9..137B,Bottino:2004qi} 
and coming from the angular direction $\psi$ is given by:
\begin{equation}
\Phi_\gamma(E_\gamma, \psi) = \frac{1}{4\pi} \frac{\langle\sigma v\rangle}{m_\chi^2} \frac{dN_\gamma}{d E_{\gamma}}
\frac{1}{2}I(\psi) \, ,
\label{eq:flux_gamma}
\end{equation}
where {\sigmav} is the annihilation cross section times the relative
velocity mediated over the galactic velocity distribution function, and
$dN_\gamma/d E_{\gamma}$ is the energy spectrum of $\gamma$-rays
originated from a single DM pair annihilation. 
In particular, we may identify WIMP candidates with 
neutralinos in the Minimal Supersymmetric Standard Model 
(see Ref. \cite{Bottino:2004qi} and refs. therein).
\\
%%%% dN/dE
The photon spectrum in the continuum originates 
 from the production of fermions, gauge bosons, Higgs bosons, and gluons from
 the annihilation of WIMP pairs. The spectra $dN_\gamma/d E_{\gamma}$ from 
DM final states into $b\bar{b}$,  $\mu^+\mu^-$ and  $\tau^+\tau^-$
have been taken from Refs. \cite{2004PhRvD..70j3529F,2011PhRvD..83h3507C}. 
The extrapolation down to m$_\chi$=10 GeV seems guaranteed 
within 10\% of uncertainty for all the 
annihilation channels \citep{lineros_priv} (a more 
careful derivation being beyond the scope of the paper). 
\\
The quantity $I(\psi)$ is the integral performed along the l.o.s. 
 of the squared DM density distribution:
\begin{equation}
 \label{Ipsi} 
 I(\psi) = \int_{l.o.s.}\rho^{2}(r(\lambda, \psi))d\lambda\,.
\end{equation} 
with $\psi$ being the angle between the l.o.s. and the direction pointing toward the 
galactic center (GC) and defined in function of the galactic coordinates so that
$\cos\psi = \cos b\cos l$. 
When comparing with experimental data, Eq. (\ref{Ipsi}) must be
averaged over the telescope observing solid angle, 
$\Delta\Omega$: 
 \begin{equation}
I_{\Delta\Omega} = \frac{1}{\Delta\Omega}\int_{\Delta\Omega}I(\psi(b, l))d\Omega\,. 
\end{equation}
The integral of the squared DM density over the line-of-sight
depends from the choice on $\rho(r)$. When including the galactic center in 
the integration (Eq. (\ref{Ipsi})), different DM distributions may lead to 
very different results for $I(\psi)$. However, since our analysis is applied to 
high latitude regions, the various descriptions for $\rho(r)$ point 
to very similar values for $I(\psi)$. We neglect any  clumpiness effects
 and assume a smooth distribution of DM in the galactic halo.
\begin{table} 
\centering 
\begin{tabular}{|l|c|c|c|c|} 
\hline Halo profile         & Isothermal   & NFW  \cite{1996ApJ...462..563N}& Einasto \cite{2004MNRAS.349.1039N}    \\ 
&a = 3.5 kpc         &a = 25 kpc  & $\alpha = 0.142$ \\ 
& & $r_{c} = 0.01$ pc & $r_{-2} = 26.4$ pc \\ 
& & & $\rho_{-2} = 0.05$ GeV cm$^{-3}$ \\ 
\hline $|b|>10^{\rm o}$ & 2.389 & 2.400  & 2.833\\  
\hline $10^{\rm o}<|b|<20^{\rm o}$  & 4.020  & 4.166 & 5.752\\
\hline $|b|>60^{\rm o}$  & 1.226  & 1.283 & 1.232\\
\hline \end{tabular} 
\caption[Values for $I_{\Delta\Omega}$ in units of GeV$^{2}$cm$^{-6}$kpc]
{Values for $I_{\Delta\Omega}$ in units
of GeV$^{2}$cm$^{-6}$kpc. For all these profiles $\rho_{l} = 0.4$ GeV cm$^{-3}$ , $R_{Sun} = 8.2$ kpc.} 
\label{tab:valuesIntegrationCode} 
\end{table} 
The results for  $I_{\Delta\Omega}$  for
different DM density distributions and observational regions are reported
 in Table \ref{tab:valuesIntegrationCode}. All the DM profiles 
provide very similar results for latitudes well above the galactic plane. 
Hereafter, the results will be provided for the cored isothermal density profile.

%%%%%%%%%%%%%%%%%%%%%%%%%%%%%%%%%%%%%%%%%%%%%%%%%%%%%%%555
\subsection{Results  on annihilation cross section}
\label{sec:upperbounds}
\begin{figure}[!h] \centering
\includegraphics[scale=0.5]{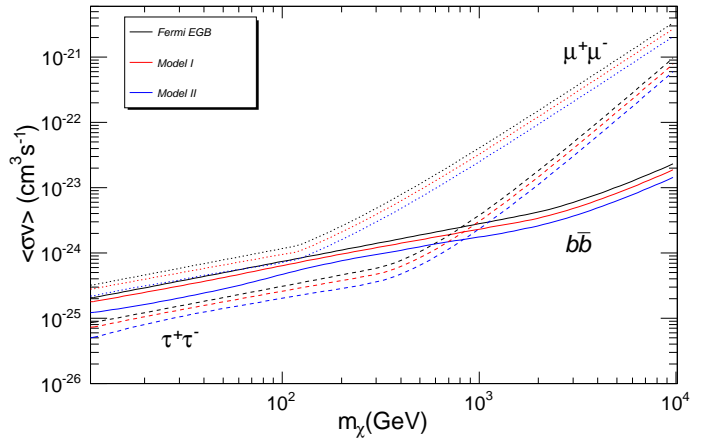}  
\caption{Upper bounds
on {\sigmav} from $\gamma$-ray in the high latitude galactic halo, 
as a function of the DM mass. 
From top to bottom, solid lines refer to 90\% C.L.
 limits  from the comparison with  Fermi-LAT EGB   (black lines), 
Model I (red lines), Model II (blue lines) (see text for details). 
Dotted, solid and dashed lines correspond 
DM annihilation into $\mu^+\mu^-$, $b\bar{b}$, $\tau^+\tau^-$, respectively.} 
\label{fig:upper_bounds}  
\end{figure}
 In this Section we
derive upper bounds  at 90\% C.L. on the WIMP annihilation cross section 
from the   $\gamma$-ray Fermi-LAT EGB and the EGB residual 
fluxes identified as Model I and II in Sect. \ref{sec_residui}.

For the Fermi-LAT EGB \cite{LATIsotropicSpectrum}, 
the upper bounds  at 90\% C.L. 
on {\sigmav}  are obtained by requiring that the DM signal 
calculated according to Eq. \ref{eq:flux_gamma} does not 
exceed the measured flux plus 1.28$\sigma$ 
(one-sided upper limit on the {\sigmav} parameter). 
The corresponding constraints are plotted as black lines in Fig. 
\ref{fig:upper_bounds}. 
From the same data  we have then subtracted the 
unresolved blazars and MSPs minimal contribution, 
as described at length in Sect. \ref{sec_residui},
 and derived the upper bounds on 
{\sigmav} corresponding to Model I (red lines).  
Similarly, upper bounds for the Model II EGB are obtained 
from the further subtraction of the minimal 
flux from star forming galaxies and UHECRs (blue lines). 
In Fig. \ref{fig:upper_bounds} we display the conservative upper bounds
on the thermal annihilation DM cross section at 90\% C.L., 
derived within the previous assumptions. 
From top to bottom, each bunch of lines refer to the limits on {\sigmav}, 
arising from the comparison with the Fermi-LAT 90\% C.L. EGB,
Model I  and Model II.
Dotted, solid and dashed lines correspond 
DM annihilation into $\mu^+\mu^-$, $b\bar{b}$, $\tau^+\tau^-$, respectively.
Given the scaling of the DM flux $\propto m_\chi^{-2}$,  constraints on {\sigmav} 
increase with the mass and span about two orders  of magnitude in the considered mass interval. 
It is evident from Fig. \ref{fig:upper_bounds} that the subtraction of the minimal 
amount of $\gamma$-rays from unresolved sources lowers the limits on {\sigmav} by at least 50\%. 
The Fermi-LAT data for the EGB are available also for latitudes $10^{\rm o}<|b|<20^{\rm o}$ and 
$|b|>60^{\rm o}$ \cite{LATIsotropicSpectrum}. The flux in Eq. (\ref{eq:flux_gamma}) 
changes for the mere normalization factors
given in Table \ref{tab:valuesIntegrationCode}. 
However, given the intensity of the measured fluxes our 
upper limits do not change if derived for the other high latitude regions. 

 Given the theoretical uncertainties affecting
the DM content and the astrophysical backgrounds, 
the results in Fig. \ref{fig:upper_bounds} are of the same order of 
magnitude or lower than the bounds 
on {\sigmav}
from cosmological DM \cite{2010JCAP...04..014A},
 from the galactic center 
\cite{Abramowski:2011hc}, or from inverse Compton processes 
evaluated from $\gamma$-rays in different portions of the sky \citep{Cirelli:2009dv}. 
Very recently (during the review process of the 
present paper) the Fermi-LAT collaboration performed 
a combined analysis on ten Milky Way satellite galaxies
\citep{fermi_dwarfs_2011}, corroborated by the analysis in Ref. 
\citep{koushiappas}. 
The absence of DM signals from
these objects leads to upper limits on  {\sigmav} which are 
close to $10^{-26}$ for masses about 10 GeV and $10^{-24}$ for 
$m_\chi$=1 TeV. These bounds are close to the ones established in 
the present work for the high mass side, and stronger 
for the low mass range. The two results, given 
unavoidable modelling in the extraction of the upper bounds, 
strengthen each other in disfavouring a DM candidate 
with an annihilation cross section much higher than the 
electroweak reference value $3 \cdot 10^{-26}$ cm$^3$/s 
for very low WIMP masses. We make notice that 
the EGB spectra we have obtained in Model I and II could be 
further reduced, whether by the subtraction of additional components or 
by increasing the predictions for each contribution,
 set at the minimum  in the present work.
A smaller $\gamma$-ray flux at high latitudes 
could therefore be as powerful as the 
measurements from the dwarf spheroidal galaxies. 
\\
We emphasize that our limits are almost model
independent: little dependence on the  DM distribution, being at high latitudes, 
and mild differences due to final states. 
\\ Our limits are
{\it conservative}: it is very unlikely that a higher {\sigmav}  
be compatible with Fermi-LAT EGB. Similarly, 
our upper limits could be lowered only
with  assumptions on non-homogeneous DM distributions or, of course, 
comparing to a smaller EGB residual.

%%%%%%%%%%%%%%%%%%%%%%%%%%%%%%%%%%%%%%%%%%%%%%%%%%%%%%%%%%%%%%%%%%% 
\subsection{Bounds on the Sommerfeld enhancement for {\sigmav}} 
\label{sect:Sommerfeld}
Recent claims on the excess of CR positrons \cite{2009Natur.458..607A} have  
stimulated the interpretation of data in terms of annihilating DM with fairly large 
annihilation cross sections of the order of $10^{-23}-10^{-22}$ cm$^3$/s. These numbers
are at least three orders of magnitude larger than the value indicated by observations
of the DM abundance due to thermal production. 
One way to boost the annihilation cross section is through the Sommerfeld effect 
\cite{somm,2009PhRvD..79a5014A,Lattanzi,hisano,iengo, Zavala}, generically due to an attractive force acting between two particles,
$i.e.$ a Yukawa or a gauge interaction. 
In the case of DM particles, the main effect of such an attractive force would be to enhance {\sigmav}
by a factor proportional to $1/\beta = c/v$, where $v$ is the velocity of the DM particle
($1/v$ enhancement). The net result on the annihilation cross section writes as  
{\sigmav} = S {\sigmav}$_{0}$, where S sizes the Sommerfeld enhancement of the annihilation amplitude.
We have evaluated the Sommerfeld enhancement S
using the approximation of the Yukawa potential by the Hulthen potential, for which an analytic
solution is possible~\cite{2010JPhG...37j5009C,Feng2} (and checked that the solution coincides with the numerical one). 
The \som factor  behaves as $1/v$  and for very small velocities it saturates to
constant values.  Given $\alpha$  the coupling constant and  $m_\phi$ the mass of the new force 
carrier, if the quantity $m_\phi/m_\chi \cdot \alpha$ is close to the values that make the Yukawa 
potential have zero-energy bound states, the enhancement is much larger; indeed, the enhanced cross 
section shows resonances at $ m_\chi = \frac{4 m_\phi n^2}{\alpha}  ~ (\rm{n} = 1,2,3...)$, 
which grow as  $1/v^2$, up to the point where they get cut off by finite width effects.

\begin{figure}[!h] \centering
\includegraphics[scale=0.45]{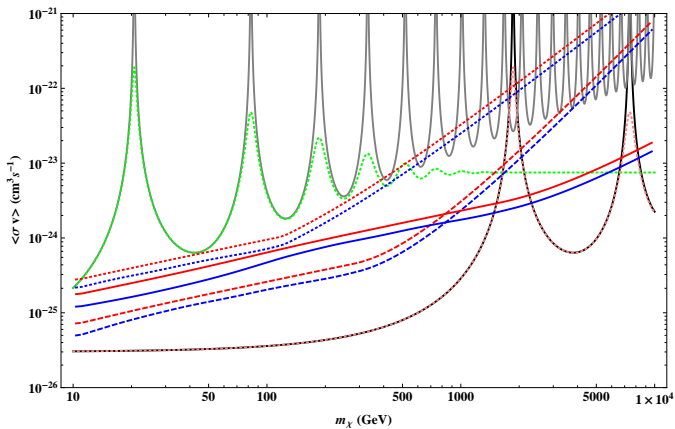}  
\caption{Sommerfeld enhancement of the annihilation cross section as a function of the DM mass, 
for $\alpha = \frac{1}{4\pi}$. Solid curves are for $\beta =10 ^{-8}$, dotted ones 
for $\beta =10 ^{-3}$. 
The upper (lower) resonant curve is obtained for a force carrier of mass $m_{\phi} = 1$ GeV (90 GeV). 
The upper (lower) dotted, solid and dashed curves  correspond to the upper bounds for EGB Model I (Model II)
derived from WIMPs annihilating in the high latitude galactic halo in $\mu^+\mu^-$, $b\bar{b}$, $\tau^+\tau^-$, 
respectively (see Fig. \ref{fig:upper_bounds}.)}
\label{fig:somm_sigmav}  
\end{figure}

In Fig. \ref{fig:somm_sigmav}  we show the Sommerfeld 
enhanced cross sections for $\alpha = \frac{1}{4\pi}$, $\beta =10 ^{-8}$ and a force carrier 
of mass $m_{\phi} = 1$ GeV (upper
curve) and  $m_{\phi} = 90$ GeV (lower curve).  We over-impose the upper bounds obtained in 
the previous Section from the residual EGB Model I and Model II and already displayed in Fig. 
\ref{fig:upper_bounds}. Our results 
show that a Sommerfeld enhancement due to a force carrier of  $m_{\phi} < 1$ GeV 
($\alpha = \frac{1}{4\pi}$)  is strongly excluded by Model I and II for the Fermi-LAT EGB data. 
For a massive force carrier (90 GeV) only the resonant peaks above the TeV mass are excluded. 
The result holds for  $\beta =10 ^{-8}$  up to $\beta =10 ^{-3}$.
Comparable constraints have been obtained in 
\cite{2009PhRvD..80b3505G,2011PhRvD..84b7302G} through
the analysis of perturbations to the CMB angular power spectrum

Therefore, high latitude $\gamma$-ray observations interpreted as due to DM annihilation in the Milky Way  
halo bound the Sommerfeld enhancement of the annihilation cross section to a factor of 3-10-50-200 for
 $m_\chi$=10-100-1000-5000 GeV, respectively. In case a Yukawa-like potential describes this
non-relativistic quantum effect, a force carrier heavier than 1 GeV is definitely required. 
%%%%%%%%%%%% 

%%%%%%%%%%%%%%%%%%%%%%%%%%%%%%%%%%%%%%%%%%%%%%%%%%
\subsection{Bounds from the high-redshift protohalos}
A possible way to boost the annihilation rate is to modify the 
particle theory and make the ansatz that the annihilation
cross section depends on the inverse of the velocity \cite{Robertson:2009bh}.
A boosted production of $\gamma$-rays in models with 
{\sigmav}$ \propto 1/v$ has been 
proposed for the first bound objects formed 
in the early phases of the universe \cite{proto,PK}. 
After the matter-radiation equality is reached, DM perturbations start growing via gravitational
instability and form the first bound protohalos at a redshift of about 140. 
 The birth of these protohalos depends on the properties of the DM particles, 
since they are responsible of the primordial inhomogeneities. 
%%%%%%%%%
The complete decoupling of DM particles from the thermal bath happens later with
respect to the freeze-out temperature $T_{f}$, at a temperature of kinetic
decoupling $T_{kd}$, because scattering events with SM particles keep the WIMPs close
to thermal equilibrium. For $T \leq T_{kd}$,
 free-streaming  and acoustic oscillations
 compete to  damp the power spectrum of matter
density fluctuations, which sets in turn the mass $M_{c}$ of primordial DM structures
\cite{Green:2003un,2005PhRvD..71j3520L,Hofmann:2001bi,Chen:2001jz}.
For very small values of $M_{c}$, the details of the QCD phase transition could
further slightly damp the actual cutoff mass \cite{1999PhRvD..59d3517S}.
In general, however, both the thermal and kinematic decouplings from cosmic plasma
are heavily linked to the WIMP nature and interactions with SM particles.  
%The complete decoupling of DM particles 
%from the thermal bath happens therefore quite later with respect to the 
%freeze-out temperature $T_{f}$, 
%at a temperature  of kinetic decoupling $T_{kd}$ which sets the distance scale 
%at which linear density perturbations in the DM distributions are wiped off. 
%This small-scale cutoff in the matter power spectrum sets in turn the mass $M_{c}$  of the
%primordial DM structures. Since the $T_{kd}$ establishes the link between the first halos' 
%mass and the DM nature, it is heavily depending on the WIMP model and on assumptions 
%about the WIMP interactions. 
%Notably, the DM particle cross section for scattering off
%Standard Model particles, sets the time of both its thermal and kinematic
% decoupling from cosmic plasma.
% 
The velocity dispersion of the first protohalos that collapse at redshift $z_C$ 
is estimated to be very small ($\beta\sim 10^{-8}$) \cite{PK}. 
Therefore, models for {\sigmav} depending on the inverse of $v$ predict a 
boosted flux of DM annihilation products.  
The photons arising from WIMP annihilations in very early 
halos can freely propagate with their energy red-shifting and 
reach the Earth in the range $\sim $ keV - TeV, while photons emitted out of this 
transparency window  are absorbed by the intergalactic medium. 
 
\begin{figure}[!h] \centering
\includegraphics[scale=0.45]{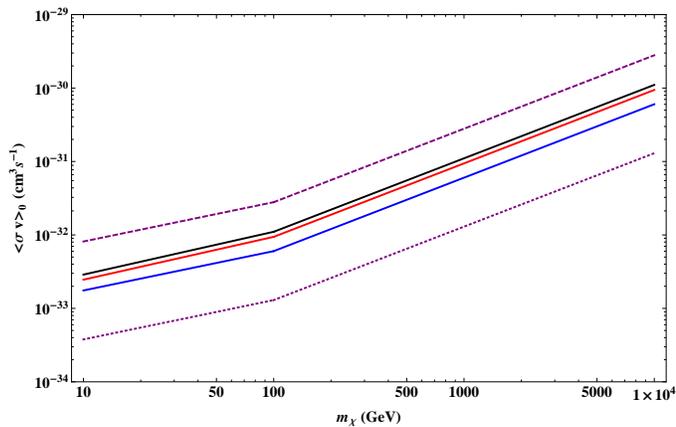}  
\caption{Bounds on {\sigmav}$_{0}$ from Eq. (\ref{eq:sigmav_1su_v}), as a function of the DM mass.
The central three bounds are obtained for $M_c = M_\oplus$, and from 
Eqs. (\ref{eq:rho_gamma_fermilat}) (black line), (\ref{eq:rho_gamma_modelI}) (red line) and 
(\ref{eq:rho_gamma_modelII}) (blue line) respectively, from top to bottom.
The upper (lower) purple lines are derived from  Eq. (\ref{eq:rho_gamma_modelII}) for Model II EGB
and $M_c = 10^{2} M_\oplus (10^{-2} M_\oplus$).
} 
\label{fig:bounds_sigmav_0}  
\end{figure}

The $1/v$ enhancement of the annihilation cross section may be simply parameterized by writing  \cite{PK}:
\begin{equation}
  \langle\sigma v\rangle=  \langle\sigma v\rangle_{0} \frac{ c}{v} \;\; {\rm cm^3/s}. 
  \label{eq:sigmav_1su_v}
\end{equation}  
\\
The energy density in photons today from WIMP annihilation in the primordial halos can be theoretically predicted by:
\begin{equation}
 \rho_{\gamma} = 5.28 \cdot 10^{6}\left(\frac{M_c}{M_\oplus}\right)^{-1/3} \langle\sigma v\rangle_{0}
B_{2.6}\left(\frac{m_\chi}{\rm{TeV}}\right)^{-1} \rm{GeV} \rm{cm}^{-3}, 
\label{eq:fotondensity1} 
\end{equation}
where the cosmological boost factor B, normalized to 2.6 ($B_{2.6}=B/2.6$)
 takes into account that the DM is distributed according to a Navarro-Frenk-White (NFW) density profile with the lowest
 concentration parameter \cite{PK}.
Eq. (\ref{eq:fotondensity1}) can be compared with the experimental photon density inferred for the 
Fermi-LAT EGB \cite{LATIsotropicSpectrum} and for our two EGB models derived in Sect. \ref{sec_residui}, which is obtained 
by integrating the photon flux on the Fermi-LAT energy range (100 MeV - 100 GeV). We obtain:
\begin{equation}
 \rho_{\gamma}  \simeq  6.62 \cdot 10^{-16}\left(\frac{E_{\gamma}}{\rm{GeV}}\right)^{-0.41}  \rm{GeV} \;\; \rm{cm}^{-3} \;\;  {\rm (Fermi-LAT)} 
 \label{eq:rho_gamma_fermilat}
\end{equation} 
 
 \begin{equation}
\rho_{\gamma}  \simeq  5.65 \cdot 10^{-16} \left(\frac{E_{\gamma}}{\rm{GeV}}\right)^{-0.41}  
\rm{GeV} \;\; \rm{cm}^{-3}  \;\;{\rm (Model \ I)} 
\label{eq:rho_gamma_modelI}
\end{equation}

\begin{eqnarray}
 \rho_{\gamma}  \simeq 4.5\cdot10^{-16}\left(\frac{E_{\gamma}}{\rm{GeV}}\right)^{-0.46} 
 \rm{GeV} \;\; \rm{cm}^{-3} \\ 
 {\rm (Model \ II, E_{\gamma}>8 {\rm \ GeV})} \nonumber
 \label{eq:rho_gamma_modelII}
\end{eqnarray}
We constrain {\sigmav}$_{0}$ by comparison of the theoretical expression 
\ref{eq:fotondensity1} with the experimental $\gamma$-ray density.
The results are displayed in Fig. \ref{fig:bounds_sigmav_0} as a function of the WIMP mass. 
The three central lines bound the 
{\sigmav}$_{0}$ (Eq. (\ref{eq:sigmav_1su_v})) parameter from Fermi-LAT photon density given in Eq. 
(\ref{eq:rho_gamma_fermilat}), 
Eq. (\ref{eq:rho_gamma_modelI})  and Eq. (\ref{eq:rho_gamma_modelII}) respectively, from top to bottom, 
when $M_c = M_\oplus$. The upper (dotted) and the lower (dashed)  bounds
are derived for Model II when  $M_c = 10^{2} M_\oplus$ (upper) and  $10^{-2} M_\oplus$ (lower). 
The bounds on {\sigmav}$_{0}$ are strong: for WIMP masses below 100 GeV it is forced to 
be $<10^{-33}$ cm$^3$/s. Upper bounds grow to $<10^{-32}$ cm$^3$/s  for $m_\chi\simeq$ 1 TeV
and sets to $<10^{-31}$ cm$^3$/s at 10 TeV. 
We make notice that they are more stringent than limits 
obtained from primordial 
light elements abundance and CMB anisotropies \cite{Hisano:2011dc}
and significantly improve the bounds of Ref. \cite{PK}.
\\
The 1/v behaviour of the {\sigmav} may be identified with the Sommerfeld 
effect for velocities $\beta\gg(m_\phi/m_\chi)^{1/2}$ \cite{Lattanzi}.
For lower velocities, as are the ones typical for protohalos, 
the series of resonances appears (see Sect. \ref{sect:Sommerfeld})
and the Sommerfeld enhancement S behaves as $1/v^2$ close to the peaks. 
In this case, the upper bounds on the annihilation cross section
may be obtained by rescaling {\sigmav}={\sigmav}$_0$S with a factor $1/\beta \cdot m_\phi/m_\chi$.
From Eqs. \ref{eq:fotondensity1} - \ref{eq:rho_gamma_modelII} 
it is straightforward to notice that the bounds on a Sommerfeld enhanced {\sigmav}
derived from a overproduction of $\gamma$-rays in protohalos, are much weaker 
than the ones imposed by annihilation in the high-latitude galactic halo.

%%%%%%%%%%%%%%%%%%%%%%%%%%%%%%%%%%%%%%%%%%%%%%%%%%%%%%%%%%%%%%%%%%%%%%%%%%
 \section{Conclusions} 
The $\gamma$-ray EGB measured by Fermi-LAT \cite{LATIsotropicSpectrum} likely 
includes contributions from galactic and extragalactic $unresolved$ sources. 
We have explored possible non-negligible diffuse contributions from 
unresolved blazars, MSPs, star-forming galaxies and UHECRs. 
Lead by a conservative attitude, we have considered the minimal contribution
for all sources, and neglected  those objects whose high latitude flux is not
 excluded to be less than 1\% of Fermi-LAT EGB.
Two residual EGB fluxes have been derived by subtraction of the 
additional fluxes from the Fermi-LAT EGB: Model I is obtained after the subtraction 
of unresolved BL Lacs, FSRQs and galactic MSPs, while Model II 
is the residual flux after the further subtraction of 
star-forming galaxies and UHECRs.  
\\
From our new residual EGB fluxes, we have set upper limits 
on the DM annihilation cross section into $\gamma$-rays. 
A conservative upper bound on {\sigmav} is derived by assuming 
that the Model I and II EGB are entirely due to WIMPs pair-annihilating 
in the halo of our Galaxy. Values for {\sigmav} $ \gsim 10^{-25}$ cm$^3$/s are strongly excluded for 
$m_\chi \simeq$ 10 GeV, while for $m_\chi \simeq$ 100 GeV (1 TeV) the annihilation rate 
is bounded to $3 \cdot 10^{-25}$ cm$^3$/s ($10^{-24}$ cm$^3$/s). This results holds for 
DM annihilating into $b\bar{b}$. Stronger limits below $m_\chi$= 1 TeV 
are derived for annihilation into the leptonic $\tau$ annihilating channel, while for the $\mu$ channel 
the limits are close to the $b\bar{b}$ below  $m_\chi$= 100 GeV, and  weaker above this mass. 
Annihilation into leptons is therefore excluded at a level which strongly 
disfavours the interpretation of cosmic positron fraction data in terms of 
leptophilic DM with small cosmological boost factors. 
The latter boost factors are in turns strongly limited by antiproton data \cite{2009PhRvL.102g1301D}.
\\
The bounds on {\sigmav} have been interpreted in terms of Sommerfeld enhancement 
of the annihilation cross section. 
A Sommerfeld enhancement due to a force carrier of  $m_{\phi} < 1$ GeV 
($\alpha = \frac{1}{4\pi}$)  is strongly excluded by Model I and II for the Fermi-LAT EGB data. 
For a massive force carrier (90 GeV) only the resonant peaks above the TeV mass are excluded. 
High latitude $\gamma$-ray observations interpreted as due to DM annihilation in the Milky Way  
halo bound the Sommerfeld enhancement of the annihilation cross section to a factor of 3-10-50-200 for
 $m_\chi$=10-100-1000-5000 GeV, respectively, and in case an annihilation into light quarks occurs. 
For  $m_\chi \lsim$ 6-700 GeV these limits are reduced by a factor of few for the pure $\tau^+\tau^-$
annihilation channel.  In case a Yukawa-like potential describes this
non-relativistic quantum effect, a force carrier heavier than 1 GeV is definitely required. 
\\
Finally, we have explored the possibility that the residual $\gamma$-ray EGB is entirely due to 
cosmological annihilation of DM in protohalos at high redshift. 
Within the hypothesis that {\sigmav} is inversely proportional to the WIMP velocity, 
very severe limits are derived for the velocity-independent part of the annihilation cross section, 
depending on the protohalo mass.

%%%%%%%%%%%%%%%%%%%%%%%%%%%%%%%%%%%%%%%%%%%%%%%%%%%%%%%%%%%%%%%%%%%%%%%%%%
%%%%%%%%%%%%%%%%%%%%%%%%%%%%%%%%%%%%%%%%%%%%%%%%%%%%%%%%%%%%%%%%%%%%%%%%%%

\begin{acknowledgments}
We warmly thank M. Ajello and L. Latronico for helpful comments on sources in the Fermi catalog. We are grateful to T. Bringmann, N. Fornengo and R. Lineros for fruitful discussions and suggestions.
F.C. thanks Vlasios Vasileiou and the Astrophysics Science Division of the NASA's Goddard Space Flight Center, 
where part of this work was done within the ISSNAF - INAF Internship Program 2010. 
This work was supported by the EU grant UNILHC PITN-GA-2009-237920, by
the Spanish MICINN under grants FPA2008-00319/FPA, by the MULTIDARK
Consolider CSD2009-00064, by Prometeo/2009/091. F.C. acknowledges support from the German
Research Foundation (DFG) through grant BR 3954/1-1.
\end{acknowledgments}

\bibliography{draftEGB}

%%%%%%%%%%%%%%%%%%%%%%%%%%%%%%%%%%%%%%%%%%%%%%%%%%%%%%%%%%%%%%%%%%%%% 
\end{document}